\documentclass[runningheads,fleqn]{cl2emult}

\usepackage{graphicx} 
\usepackage{subeqnar} 
\usepackage{phys}     

\begin{document}

\title*{Monte Carlo Simulation of Spin Models with Long-Range Interactions}

\toctitle{Monte Carlo Simulation of Spin Models with Long-Range Interactions}

\titlerunning{Monte Carlo Simulation of Spin Models with Long-Range
Interactions}

\author{Erik Luijten}

\authorrunning{Erik Luijten}

\institute{Max-Planck-Institut f\"ur Polymerforschung, D-55021 Mainz,
           Germany \\ 
           Institut f\"ur Physik, Johannes Gutenberg-Universit\"at,
           D-55099 Mainz, Germany}

\maketitle

\begin{abstract}
  An efficient Monte Carlo algorithm for the simulation of spin models with
  long-range interactions is discussed. Its central feature is that the number
  of operations required to flip a spin is independent of the number of
  interactions between this spin and the other spins in the system. In
  addition, critical slowing down is strongly suppressed. In order to
  illustrate the range of applicability of the algorithm, two specific examples
  are presented. First, some aspects of the Kosterlitz--Thouless transition in
  the one-dimensional Ising chain with inverse-square interactions are
  calculated.  Secondly, the crossover from Ising-like to classical critical
  behavior in two-dimensional systems is studied for several different
  interaction profiles.
\end{abstract}

\section{Introduction}
For a long time, systems with long-range interactions have posed a great
challenge to the application of Monte Carlo~(MC) methods. Due to the large
number of interactions that have to be taken into account, simulations are very
time consuming and hence restricted to relatively small system sizes.
Alternatively, the interaction is truncated beyond a certain distance, which
may introduce severe errors in the calculation. Thus, it is desirable to have
an algorithm at one's disposal that allows the efficient simulation of systems
with long-range interactions without making any approximation except for the
inherent statistical errors. A few years ago, an algorithm fulfilling these
requirements was indeed introduced for the simulation of spin
models~\cite{lr-alg}. It is based on the cluster method introduced by Swendsen
and Wang~\cite{swendsen87} and has the crucial property that the number of
operations required to add a single spin to a cluster is \emph{independent} of
the number of interactions of this spin with other spins in the system. Thus,
the numerical effort to simulate ferromagnetic long-range models becomes
comparable to that for short-range models.  Close to the critical point the
application of the algorithm becomes particularly profitable, since there one
also benefits from the inherent reduction in critical slowing down in cluster
algorithms.  It is the purpose of the present paper to discuss this algorithm
in some more detail and to illustrate its power by means of some selected
applications.

\section{Description of the Algorithm}

\subsection{Conventional Wolff Cluster Algorithm}
To set the stage, we first briefly discuss the standard cluster algorithm for a
system with short-range interactions.  For simplicity, we will employ a
formulation in terms of the Wolff cluster algorithm~\cite{wolff89} rather than
that of Swendsen and Wang~(SW).  It should be noted, however, that the
modifications required for the simulation of long-range interactions pertain to
the cluster-formation process only, and hence can equally well be combined with
the SW variant. Since a detailed derivation of the Wolff algorithm can be found
in numerous references, we restrict ourselves here to a schematic outline.
Consider a $d$-dimensional lattice structure, where on each lattice site~$i$ a
spin~$s_i$ is placed, which can take the values~$\pm 1$ and which has a
ferromagnetic coupling~$K$ with its nearest neighbors. Thus, the system is
described by the following Hamiltonian,
\begin{equation}
\beta \mathcal{H} = - K \sum_{i < j} s_i s_j \;,
\end{equation}
where the sum runs over all pairs of nearest neighbors. At each Monte Carlo
step, a cluster of spins with identical sign is flipped. The nonlocal character
of this operation leads to a rapid decay of correlations. The crucial step is
the identification of a cluster:
\begin{enumerate}
\item Randomly choose a spin~$s_i$ from the lattice.
\item Consider each spin that interacts with spin~$s_i$. It is added to the
  cluster with a probability $p = 1 - \exp(-2K)$, provided that it has the same
  sign as $s_i$.
\item Repeat step~2 in turn for each spin that is newly added to the cluster,
  where one now considers all spins that interact with this spin, rather than
  with~$s_i$. This is iterated until all neighbors of all spins in the cluster
  have been considered for inclusion.
\end{enumerate}
This cluster-building process is based on the Kasteleyn--Fortuin
mapping~\cite{kasteleyn69,fortuin72} of the Potts model on a bond-percolation
model. This mapping allows a generalization of the outlined procedure to
systems with different interaction strengths~$K_j$. In the second step, a spin
is simply added with a probability $p_j = 1 - \exp(-2K_j)$ if the interaction
strength with the spin~$s_i$ is equal to~$K_j$. As a side note, it is remarked
that this implies that a cluster now not necessarily can be visualized as a
contiguous collection of spins.

Although one can thus easily devise a cluster algorithm for systems with
long-range interactions, its efficiency rapidly decreases with an increasing
number of interactions. Indeed, the typical value for the coupling~$K$ in such
a system will be relatively small and hence each single spin that is considered
for inclusion in the cluster will be added only with a small probability. So,
a lot of operations are required to add a single spin to the cluster.

\subsection{Efficient Construction of Clusters}
In order to illustrate our approach, we take the example of a
one-dimensional chain with a spin--spin interaction that decays as a
power-law, $K_{ij}=f r_{ij}^{-1-\sigma}$. The cluster-building process
starts with a spin on a randomly chosen site~$i$ and all other spins
in the system are added to the cluster with a probability $p(s_i,s_j)
= \delta_{s_i s_j} p_{ij}$, where $p_{ij} = 1 -
\exp[-2f|i-j|^{-(1+\sigma)}]$. For each spin that is actually added to
the cluster, its address is also placed on the \emph{stack}. When all
spins interacting with the first one have been considered, we read a
new spin from the stack and repeat the process, until the stack is
empty. The spin from which we are currently adding spins will be
called the \emph{current spin}.  In order to avoid considering each
single spin for inclusion in the cluster, we split up the
probability~$p(s_i,s_j)$ into two parts, namely the Kronecker delta
testing whether the spins $s_i$ and $s_j$ have the same sign and the
``provisional'' probability~$p_{ij}$. This enables us to define the
concept of the \emph{cumulative probability} $C(j)$, from which we can
read off which spin is the next one to be provisionally added to the
cluster,
\begin{equation}
\label{eq:cumprob}
  C(j) \equiv \sum_{n=1}^j P(n)
\end{equation}
with
\begin{equation}
\label{eq:firstbond}
  P(n) = \left[ \prod_{m=1\vphantom{j}}^{n-1} (1-p_m) \right] p_n \;.
\end{equation}
$p_j \equiv 1-\exp(-2K_j)$ is an abbreviation for $p_{0j}$ (and $K_j \equiv
K_{0j}$), i.e., we define the origin at the position of the current spin.
$P(n)$ is the probability that in the first step $n-1$ spins are skipped and
the $n$th spin is provisionally added. Now the next spin~$j$ that is
provisionally added is determined by a (pseudo)random number $g \in [0,1
\rangle$: $j-1$ spins are skipped if $C(j-1) \leq g < C(j)$. If the $j$th spin
has indeed the same sign as the current spin then $s_j$ is added to the
cluster.  Subsequently we skip again a number of spins before another spin at a
distance $k>j$ is provisionally added. Due to the requirement $k>j$ we must
shift the function~$P$,
\begin{equation}
  \label{eq:shiftfirstbond}
   P_j(k) = \left[ \prod_{m=j+1}^{k-1} (1-p_m) \right] p_k \;,
\end{equation}
and Eq.~(\ref{eq:firstbond}) is simply a special case of
Eq.~(\ref{eq:shiftfirstbond}). The appropriate cumulative probability is now
given by a generalization of Eq.~(\ref{eq:cumprob}),
\begin{equation}
\label{eq:gen-cumprob}
 C_j(k) = \sum_{n=j+1}^{k} P_j(n) \;.
\end{equation}
By using the specific form of the probability~$p_{ij}$ one finds that
this reduces to
\begin{equation}
\label{eq:gcumprob}
 C_j(k) = 1 - \exp \left( - 2 \sum_{n=j+1}^{k} K_n \right) \;,
\end{equation}
i.e., the probability that the next spin that will be added lies at a distance
in the range $[j+1,k]$ is given by an expression that has the same form as the
original probability, in which the coupling constant is replaced by the sum of
all the couplings with the spins in this range.

We now consider two possibilities of calculating the distance~$k$ from a given
$C_j(k)$. A straightforward approach is the construction of a look-up table.
This means that we explicitly carry out the sum in~(\ref{eq:gcumprob}) for a
large number of distances $k$, up to a certain cutoff~$M$, and store the
results in a table. Then, after drawing a random number, we can derive the
corresponding spin distance from this table. In principle we need for each
value of~$j$ another look-up table containing the $C_j(k)$. This is hardly
feasible and fortunately not necessary, as follows from a comparison of
Eqs.~(\ref{eq:cumprob}) and~(\ref{eq:gen-cumprob}). Namely (assume $k>j$),
\begin{eqnarray}
 C(k) \hspace{\arraycolsep} = \hspace{\arraycolsep}
        C_0(k) &=& C(j) + \left[\prod_{i=1}^{j}(1-p_i) \right] C_j(k)
                   \nonumber \\
               &=& C(j) + \left[ 1 - C(j) \right] C_j(k)
\end{eqnarray}
or $C_j(k) = [C(k)-C(j)]/[1-C(j)]$. So we can calculate $C_j(k)$ directly
from~$C(k)$. In practice one realizes this by using the distance~$j$ of
the previous spin that was provisionally added to rescale the (new) random
number $g$ to $g' \in [C(j),1\rangle$; $g'=C(j)+[1-C(j)]g$.  Since we only
consider ferromagnetic interactions, $\lim_{j \to \infty} C(j)$ exists and is
smaller than 1, cf.\ Eq.~(\ref{eq:gcumprob}).

This method is very fast, since we have to calculate all cumulative
probabilities only once, but it has two major drawbacks. First, we can
accommodate only a limited number of spin distances in our look-up table and
must therefore devise some approximation scheme to handle the tail of the
long-range interaction, which is essential for the critical behavior in the
case of slowly decaying interactions (small~$\sigma$). This issue is addressed
below. Secondly, this method is impractical in more than one dimension, as the
number of distances for which the cumulative probability has to be calculated
rapidly increases with the dimensionality of the system (for a fixed cutoff).

For interactions that can be explicitly summed there exists a powerful
alternative.  In those cases, Eq.~(\ref{eq:gcumprob}) can be solved for~$k$,
yielding an expression for the spin distance in terms of $C_j(k)$, i.e., in
terms of the random number~$g$. Especially for the study of critical phenomena
this is often a feasible approach, since in many cases one can modify an
interaction such it can be explicitly integrated by only adding irrelevant
terms that do not affect the universal critical properties. As an example, we
consider again the power-law interaction $K_{ij}=f |i-j|^{-(1+\sigma)}$.  The
corresponding sum appearing in the right-hand side of~(\ref{eq:gcumprob}) is
(for $j=0$) the truncated Riemann zeta function, which cannot be expressed in
closed form. However, upon approximation of this sum by the integral
\begin{equation}
  \label{eq:contprob}
  \sum_{n=j}^{k} K_n = \sum_{n=j}^{k} \frac{f}{n^{1+\sigma}} \approx
  f \int_{j-\frac{1}{2}}^{k+\frac{1}{2}} dx\; x^{-(1+\sigma)}
\end{equation}
one still has an exact Monte Carlo scheme, but for an interaction
\begin{equation}
  \label{eq:continteraction}
  K(|i-j|) = f\int_{|i-j|-\frac{1}{2}}^{|i-j|+\frac{1}{2}} dx\;
  x^{-(1+\sigma)} \;.
\end{equation}
Both interactions belong to the same universality class, so that, e.g., for the
determination of critical exponents one is free to make the most convenient
choice. Of course, all nonuniversal quantities, such as the critical
temperature, will have different values. For the modified
interaction, Eq.~(\ref{eq:gcumprob}) reduces to
\begin{equation}
 C_j(k) = 1 - \exp \left[ - \frac{2f}{\sigma}
          \left( \frac{1}{j^\sigma} - \frac{1}{k^\sigma} \right) \right] \;,
\label{eq:contcumprob}
\end{equation}
where it should be noted that this is only the probability for adding a spin
that lies in a single direction. We have not written $j+\frac{1}{2}$ and
$k+\frac{1}{2}$ instead of $j$ and~$k$, because we prefer to use continuous
coordinates, which only in the last stage are rounded to a lattice site.
Equating $C_j(k)$ to the random number $g$ we find
\begin{equation}
 k = \left[ j^{-\sigma} + \frac{\sigma}{2f} \ln(1-g) \right] ^{-1/\sigma} \;.
\label{eq:bonddistance}
\end{equation}
Rescaling of the random number is no longer required: the lowest value, $g=0$,
leads to a provisionally added spin at the same distance as the previous one,
$k=j$. If $g=C_j(\infty) = 1-\exp[-(2f/\sigma)j^{-\sigma}]$ the next
provisionally added spin lies at infinity and thus $g \in
[C_j(\infty),1\rangle$ yields no spin at all. Once the coordinates of the next
provisionally added spin have been calculated by rounding to the nearest
integer coordinate, the periodic boundary conditions are applied to map this
coordinate onto a lattice site. For the following provisionally added spin, $j$
is set equal to $k$ (\emph{not} to the rounded distance!) and a new $k$ is
determined. If no spin has been added yet, $j$ is set to $\frac{1}{2}$, the
lowest possible distance. For higher dimensionalities, $d-1$ additional random
numbers are required to determine the direction in which the next spin is
added.

The above-mentioned limited size~$M$ of the look-up table can now be coped with
as well: beyond the spin distance $M$ the sum in~(\ref{eq:gcumprob}) is
approximated by an integral. Thus, if the random number~$g$ lies in the
interval $[C(M),C(\infty)\rangle$, the spin distance $k$ is determined from a
modified version of~(\ref{eq:bonddistance}), where the lower part of the
integral is replaced by an explicit sum
\begin{equation}
 k = \left[ \left(M+\frac{1}{2}\right)^{-\sigma} + \sigma
  \left( \frac{1}{2f} \ln (1-g) + \sum_{n=1}^{M}\frac{1}{n^{1+\sigma}} \right)
  \right]^{-1/\sigma} \;.
\label{eq:bonddist-1d}
\end{equation}

\subsection{Discussion}
Finally, we briefly illustrate the efficiency of the algorithm by means of an
example. If the probability $p$ were equal for each spin pair, one out of
$p^{-1}$ spins would be added to the cluster, and it would take
$\mathcal{O}(p^{-1}) \sim \mathcal{O}(L^0)$ operations per spin to update a
configuration, compared to $\mathcal{O}(L^d)$ operations \emph{per spin} for a
Metropolis algorithm.  Taking into account the decrease in critical slowing
down, we see that the efficiency of this method is typically a factor
$\mathcal{O}(L^{d+z})$ larger than the conventional Monte Carlo algorithm. In
the case of a large but finite interaction range~$R$, such as for the
two-dimensional equivalent-neighbor model discussed in Sec.~\ref{sec:ising2d},
one gains a factor~$\mathcal{O}(R^d L^z)$, provided that critical slowing down
is completely suppressed in the cluster algorithm. Far from the critical
temperature, the factor $L^z$ disappears, but one still has the advantage of a
speed increase proportional to the number of interactions per spin.

\section{Applications}

\subsection{The Inverse-Square Ising Chain}
\begin{figure}
  \includegraphics[width=\textwidth]{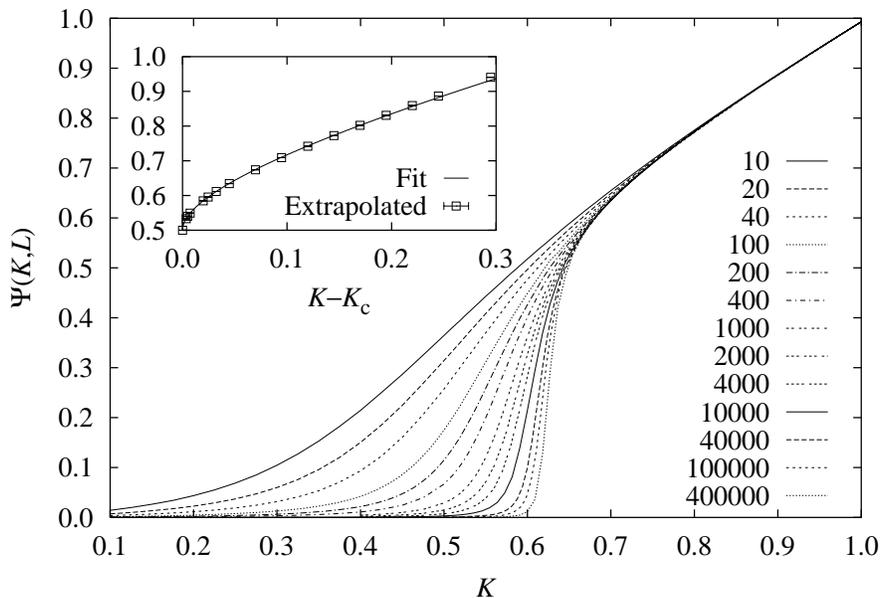}
  \caption[]{The quantity $\Psi = Km^2$ as a function of the coupling~$K$, for
  system sizes $10 \leq L \leq 400\,000$. The inset shows $\lim_{L \to \infty}
  \Psi(K,L)$ for $K > K_{\rm c}$.}
  \label{fig:psi}
\end{figure}

As a first application, we consider the Ising chain introduced before, with
spin--spin interactions that decay as $f r_{ij}^{-1-\sigma}$. Unlike the Ising
chain with short-range interactions, in which no long-range order is possible
for nonzero temperatures, this system exhibits a remarkably rich behavior.  The
interested reader is referred to Ref.~\cite{class-lr} for a review. In summary,
the system exhibits mean-field-like critical behavior for $0 < \sigma \leq
\frac{1}{2}$ and nonclassical critical behavior for $\frac{1}{2} < \sigma < 1$,
with critical exponents that depend on~$\sigma$. For $\sigma > 1$ the
interactions are essentially short-ranged. At $\sigma=1$, the so-called
inverse-square Ising model, we have a very interesting situation: the spin
chain displays a phase transition which is the one-dimensional analog of the
Kosterlitz--Thouless~(KT) transition in the two-dimensional $XY$~model.  It has
close connections to a variety of physical applications, such as the Kondo
problem, quantum tunneling in a two-state system coupled to a dissipative
environment, quark confinement, etc. Although the KT transition has received an
enormous amount of attention over the past decades, there are still open
questions. Two peculiar properties of this transition are: (1)~at the critical
temperature~$T_{\rm c}$ the order parameter exhibits a singular behavior
\emph{superposed} on a jump like one finds for a first-order transition;
(2)~the correlation length~$\xi$ and the susceptibility~$\chi$ diverge
exponentially for $T \downarrow T_{\rm c}$: $\xi = \xi_0 \exp[B_\xi / (T-T_{\rm
c})^{\nu}]$ and $\chi = \chi_0 \exp[B_\chi / (T-T_{\rm c})^{\nu}]$, with
$\nu=\frac{1}{2}$. These and related critical properties turn out to be very
difficult to verify in numerical simulations of two-dimensional models, because
the finite-size effects decay only logarithmically. The one-dimensional model,
which now can be simulated with comparable efficiency, clearly offers a great
advantage: rather than a linear system size of $\mathcal{O}(10^3)$, like for
$d=2$, one can reach system sizes $L = \mathcal{O}(10^6)$ and thus approach the
critical point much closer. Since a detailed description of the critical
properties and their numerical determination is beyond the scope of this paper,
we restrict ourselves here to a study of the quantity $\Psi \equiv Km^2$, where
$K$ is the coupling constant and $m$ the magnetization density, and its
application for the determination of~$T_{\rm c}$ (or, equivalently, its
inverse, the critical coupling~$K_{\rm c}$). It can be shown that $\Psi$ is the
analog of the spin-wave stiffness in the 2D~$XY$ model. Just like the latter
quantity is predicted to have a universal jump~$2/\pi$ at criticality, $\Psi$
is expected to have a jump of size~$\frac{1}{2}$~\cite{lr-kt}. Indeed, in
Figure~\ref{fig:psi}, where $\Psi(K,L)$ is shown for system sizes up to
$L=400\,000$, one can already clearly observe how such a jump develops with
increasing system size. The superposed square-root singularity,
\begin{equation}
\label{eq:psi-sing}
\Psi(K,\infty) = \Psi(K_{\rm c},\infty) + C\sqrt{K-K_{\rm c}} 
                 + \mathcal{O}(K-K_{\rm c}) \;,
\end{equation}
is shown in the inset.

We consider now three distinct ways to use this quantity for the determination
of~$K_{\rm c}$, which for phase transitions in this universality class is
notoriously difficult. First, one may use the predicted singular behavior of
$\Psi$ for $K>K_{\rm c}$. To this end, the finite-size data at fixed couplings
first must be extrapolated to the thermodynamic limit. By integrating the RG
equations one finds~\cite{lr-kt} that $\Psi(K,L)$ obeys a finite-size expansion
of the form
\begin{equation}
\Psi(K,L) = \Psi(K,\infty) 
            \left\{ 1 + a_1 L^{-2[\bar{\Psi}-1]} 
                      + a_2 L^{-4[\bar{\Psi}-1]}
                      + \cdots  
            \right\} \;,
\end{equation}
where $\bar{\Psi}=\Psi(K,\infty)/\Psi(K_{\rm c},\infty)$ and the ellipsis
denotes higher-order terms. The resulting estimates for $\Psi(K,\infty)$ are
then fitted to Eq.~(\ref{eq:psi-sing}), in which $\Psi(K_{\rm c},L)$ is kept
fixed at~$\frac{1}{2}$. This yielded $K_{\rm c}=0.6552~(2)$.  Secondly,
$\Psi(K,L)$ in the high-temperature regime $K<K_{\rm c}$ may be used to
estimate~$K_{\rm c}$.  Indeed, within the finite-size regime one expects two
types of corrections to scaling: corrections due to irrelevant fields, which
decay as powers of $1/\ln L$, and temperature-dependent corrections which can
be expanded in terms of $L/\xi$. A least-squares fit of the numerical data to
an expansion of the form
\begin{equation}
\label{eq:psi-ht}
\Psi(K,L) = \Psi(K_{\rm c},\infty)
            + a_1 \frac{L}{\xi} + a_2 \left(\frac{L}{\xi}\right)^2 + \cdots
            + \frac{b_1}{\ln L} + \frac{b_2}{(\ln L)^2} + \cdots 
\end{equation}
has yielded $\Psi(K_{\rm c},\infty)=0.496~(3)$ and $K_{\rm c}=0.6548~(14)$.
Fixing $\Psi(K_{\rm c},\infty)$ at the predicted value~$\frac{1}{2}$ we found
$K_{\rm c}=0.6555~(4)$.  Finally, a very straightforward but remarkably
effective approach is to fit a set of finite-size data for $\Psi(K,L)$ at
\emph{fixed} coupling to an expression of the form of Eq.~(\ref{eq:psi-ht}), in
which all temperature-dependent terms have been omitted, i.e., one
\emph{assumes} that $K=K_{\rm c}$. Although in principle such a fit should only
work at the true critical coupling, it turns out that least-squares fits of a
good quality can be obtained over a range of couplings.
\begin{table}[b]
\caption{Some estimates (in chronological order) for the critical coupling
$K_{\rm c}$ of the inverse-square Ising chain.}
\label{tab:kt-kc}
\centering
\renewcommand{\arraystretch}{1.1}
\begin{tabular}{cll}
\hline
Reference          & $K_{\rm c}$              & Method \\
\hline
\cite{nagle70}     & $\approx 0.612$          & Exact summation 
                                                + Pad\'e approximants \\
\cite{anderson71}  & $\approx 0.635$          & RG \\
\cite{doman81}     & $\approx 0.490$          & Extended mean-field approach \\
\cite{matvienko85} & $0.657~(3)$              & Series expansion \\
\cite{vigfusson86} & $\geq 0.441$             & Extended mean-field approach \\
\cite{aizenman88}  & $\geq \frac{1}{2}$       & Analytical \\
\cite{monroe90}    & $\approx 0.562$          & Coherent-anomaly method \\
\cite{wragg90}     & $0.5$                    & Variational method \\
\cite{manieri92}   & $0.590~(5)$              & Cycle expansion \\
\cite{glumac93}    & $0.615$                  & Transfer matrix \\
\cite{monroe94}    & $\geq 0.61128$           & Extended mean-field approach \\
\cite{cannas96}    & $6/\pi^2 \approx 0.6079$ & RG (conjectured to be exact) \\
This work          & $0.6552~(2)$             & MC \\
\hline
\end{tabular}
\end{table}
However, the resulting estimate of $\Psi(K,\infty)$ is a monotonously
increasing function of~$K$ and $K_{\rm c}$ can be determined from the
requirement that $\Psi(K_{\rm c},\infty)=\frac{1}{2}$, yielding $K_{\rm
c}=0.65515~(20)$.  It is rewarding that the three different methods yield
consistent estimates for the critical coupling.  Table~\ref{tab:kt-kc}
summarizes several estimates for~$K_{\rm c}$, obtained by a variety of methods.
Remarkably, most numerical results (of which only few carry an explicit
estimate of the uncertainty) suggest a critical coupling around~$0.61$. Our
estimate lies considerably higher and only agrees with the series-expansion
result of Ref.~\cite{matvienko85}. All methods that rely on an extension of
mean-field theory clearly suffer from the extremely slow convergence to the
thermodynamic limit. The exact conjecture of Ref.~\cite{cannas96} (which is,
interestingly, precisely twice as large as the mean-field result~$3/\pi^2$) has
been refuted.

\subsection{Crossover from Ising-Like to Classical Critical Behavior}
\label{sec:ising2d}
In order to illustrate that also for interactions with a large but strictly
finite range the presented algorithm offers great advantages, we consider the
so-called ``equivalent-neighbor'' model. This is a simple generalization of the
Ising model, in which each spin interacts equally strongly with all its
neighbors within a distance~$R_m$. In the limit $R_m \to \infty$ this model
becomes equivalent to the exactly-solved mean-field model, whereas all systems
with a \emph{finite} interaction range belong to the Ising universality class.
Since an exact solution for the latter case is lacking for $d=3$, it is
interesting to study the variation of critical properties as a function
of~$R_m$ (cf.\ Ref.~\cite{mr3d}). Another, experimentally very relevant,
application of this model will be discussed here. As is well known, many
thermodynamic properties show a characteristic power-law divergence upon
approach of a critical point. These powers, or critical exponents, have
universal values that only depend on a small number of \emph{global} properties
of the system under consideration. For example, binary mixtures, simple fluids,
and uniaxial ferromagnets all exhibit the same set of critical exponents as the
three-dimensional Ising model.  However, the corresponding power-law behavior
is only observed asymptotically close to the critical point, whereas at
temperatures farther away from~$T_{\rm c}$ (but still relatively close to it)
one may observe classical or mean-field-like critical behavior. This
\emph{crossover} can be explained in terms of competing fixed points of a
renormalization-group transformation and is in principle well understood.
However, unlike the critical exponents, for which accurate results have been
obtained from series expansions, renormalization-group calculations,
experiments, and numerical calculations, the precise nature of the crossover
from one universality class to another is still a point of discussion. In
particular, it is an unsettled question to what extent this crossover is
universal. There exist several field-theoretic calculations, but it has not yet
been possible to verify their correctness by means of experiments. Measurements
in the critical region are not only difficult, one also has to take great care
to make the temperature distance to the critical point not too large, since one
then would leave the critical region. As stated by the Ginzburg criterion, the
crossover is a function of $t/G$, where $t = (T-T_{\rm c})/T_{\rm c}$ is the
reduced temperature and $G$ a system-dependent parameter. Throughout the
crossover region, $t$ has to be kept small, but $t/G$ has to be varied over
several decades. The large extent of the crossover region also emphasizes its
experimental relevance: many measurements of critical exponents are actually
made within this region rather than asymptotically close to the critical point
and hence a detailed knowledge of crossover functions is required for a proper
interpretation of the data.  Since the Ginzburg parameter~$G$ is a function of
the interaction range, it is well possible to construct such crossover
functions from data obtained by means of simulations of systems with different
interaction ranges, where for each system $t$ is varied over a limited range
only. In view of the large coordination numbers [$\mathcal{O}(10^4)$] that have
to be reached within this approach, this is only feasible with an advanced
algorithm.

\begin{figure}[b]
  \includegraphics[width=\textwidth]{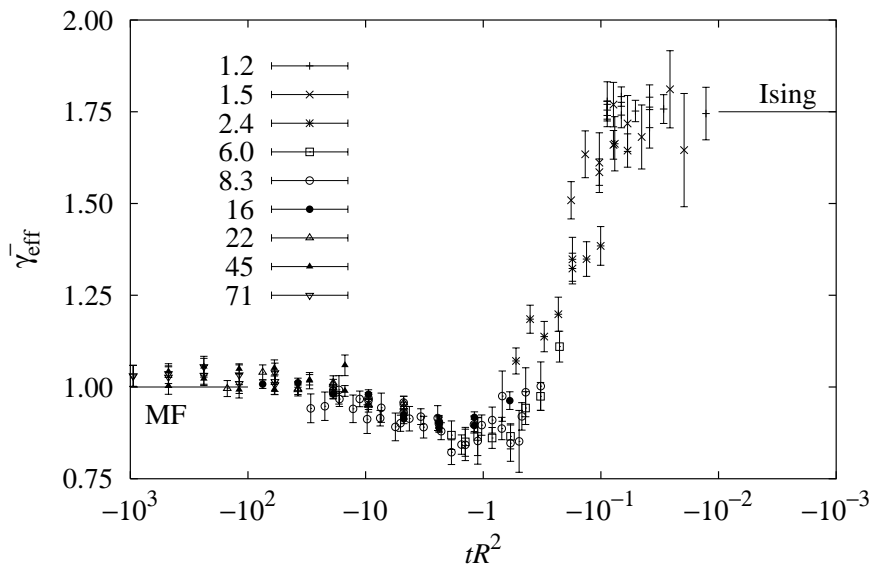}
  \caption[]{The effective susceptibility exponent $\gamma_{\rm eff}^{-}$ for
  $T<T_{\rm c}$ as a function of the crossover variable~$tR^2$.}
  \label{fig:gamma-lt-2d}
\end{figure}

\begin{figure}[t]
  \includegraphics[width=\textwidth]{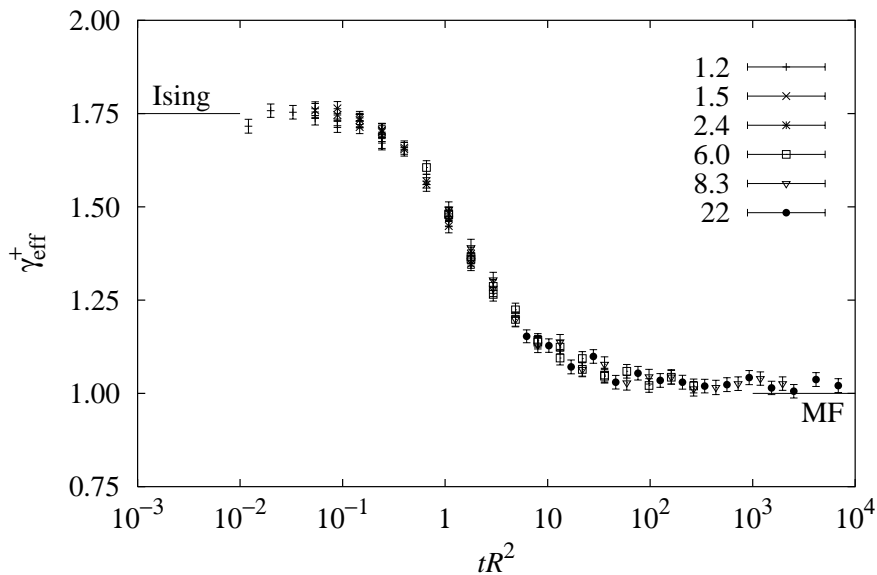}
  \caption[]{The effective susceptibility exponent $\gamma_{\rm eff}^{+}$ for
  $T>T_{\rm c}$ as a function of the crossover variable~$tR^2$.}
  \label{fig:gamma-ht-2d}
\end{figure}

We concentrate here on one specific crossover function, namely that for the
susceptibility in a two-dimensional~(2D) system, both below and above the
critical temperature (see Refs.~\cite{chicross,cross} for a more detailed
discussion of this topic). In the 2D Ising model, the susceptibility~$\chi$
diverges for $t \uparrow 0$ as $A_{\rm I}^{-}(-t)^{-7/4}$ and for $t \downarrow
0$ as $A_{\rm I}^{+}t^{-7/4}$, where the amplitudes $A_{\rm I}^{\pm}$ are known
exactly. Mean-field theory, on the other hand, predicts a susceptibility that
for $t \uparrow 0$ diverges as $1/(-2t)$ and for $t \downarrow 0$ as $1/t$.  It
is our aim to numerically determine the effective susceptibility exponent
$\gamma_{\rm eff}^{\pm} = - d \ln \chi / d \ln t$, which is expected to
interpolate smoothly between $7/4$ and~$1$.  We have carried out MC simulations
for square lattices with a maximum linear size $L=1000$ and interaction ranges
up to $R_m^2=10000$ (coordination number $z=31416$). It has been shown that the
Ginzburg parameter $G \propto R^{-2}$, where the effective interaction range
$R^2 = z^{-1} \sum_{j \neq i} |\mathbf{r}_i - \mathbf{r}_j|^2$ ($|\mathbf{r}_i
- \mathbf{r}_j| \leq R_m$) has been introduced to avoid lattice effects.  For
$t<0$ the susceptibility is calculated from the fluctuation relation $\chi' =
L^d(\langle m^2\rangle - \langle |m| \rangle^2)/k_{\rm B}T$, whereas $\chi =
L^d \langle m^2 \rangle / k_{\rm B}T$ was used for $t>0$. The exponent
$\gamma_{\rm eff}^{\pm}$ is obtained by numerical differentiation and shown as
a function of $t/G$ in Figs.~\ref{fig:gamma-lt-2d} and~\ref{fig:gamma-ht-2d},
respectively.

Several observations can be made concerning these graphs. In the first place,
one notices that both functions smoothly interpolate between the Ising exponent
and its classical counterpart. Also the expectation that the crossover takes
several decades in the parameter $t/G$ is confirmed. However, there is a
striking difference between $\gamma_{\rm eff}^{-}$ and~$\gamma_{\rm eff}^{+}$:
whereas the latter exhibits a gradual decrease from $7/4$ to~$1$ when the
distance to the critical point is increased, the former drops much more rapidly
and even goes through a minimum where $\gamma_{\rm eff} < 1$, before reaching
its asymptotic value. Interestingly, the occurrence of such a nonmonotonic
variation has been inferred from RG calculations for three-dimensional systems.
The results presented here are the first to show that, at least in the
low-temperature regime, this behavior can actually be observed in systems as
simple as the two-dimensional Ising model with an extended interaction range.

The fact that the curves in Figs.~\ref{fig:gamma-lt-2d}
and~\ref{fig:gamma-ht-2d} overlap for different values of~$R$ suggests that the
crossover functions possess a certain degree of universality. However, to what
extent one may expect such universality in a region where the correlation
length no longer diverges is still an unanswered question. It has been
suggested~\cite{anisimov95} that an additional length scale comes into play
here. In order to investigate this possibility, we have carried out simulations
of two-dimensional models that are very similar to those studied above, except
that the interaction profile has been modified~\cite{dblock}. Instead of a
constant interaction strength (block-shaped profile), a double-blocked
potential was used, where each spin interacts with a strength~$K_1$ with its
$z_1$ neighbors within a distance $r \leq R_1$ (domain~$D_1$) and with a
strength~$K_2$ with its $z_2$ neighbors within a distance $R_1 < r \leq R_2$
(domain~$D_2$). In order to create a strong asymmetry, a strength ratio
$\alpha=K_1/K_2=16$ was chosen. The outer interaction range was kept fixed at
$R_2=\sqrt{140}$, whereas the parameter~$R_1$ was varied. This allowed us to
realize two additional, qualitatively different situations: both $R_1=\sqrt{4}$
and $R_1=\sqrt{93}$ lead to an effective interaction range $R \approx
\sqrt{50}$ [the previous definition for~$R$ is now generalized to $R^2= (\alpha
\sum_{i \in D_1} r_i^2 + \sum_{i \in D_2} r_i^2)/(\alpha z_1 + z_2)$] , but in
the former case the inner domain contains only $12$ out of $436$ interaction
neighbors, compared to $292$ out of $436$ in the latter case.  This means that
the integrated coupling ratio $\alpha z_1/z_2$, which indicates the relative
contribution of the two domains to the total integrated coupling, is $0.45$ in
the first case and~$32$ in the second case. Extensive Monte Carlo simulations
have been carried out for these and several other double-blocked interaction
profiles. After a determination of the critical properties, we have calculated
the crossover curve for the susceptibility in the region $t<0$ for each of
these profiles, in order to see whether the nonmonotonicity in
Fig.~\ref{fig:gamma-lt-2d} is a peculiarity of the equivalent-neighbor model.
The critical temperature, normalized by the critical temperature of the
mean-field model, indeed displays a clear nonuniversality: for $R_1^2=93$ the
critical fluctuations turn out to be stronger suppressed than for $R_1^2=4$,
even though the effective interaction range is almost identical.
\begin{figure}
  \includegraphics[width=\textwidth]{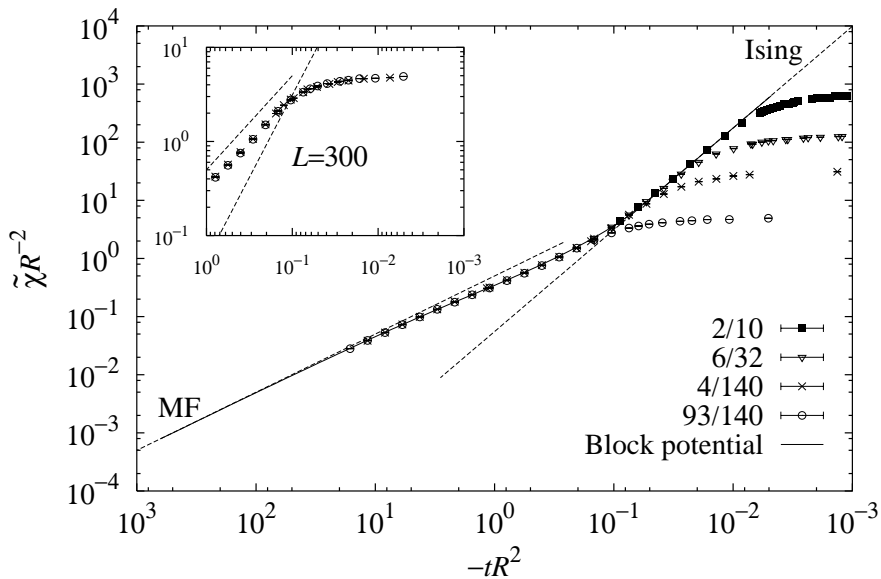}
  \caption{The crossover function for the connected susceptibility
  below~$T_{\rm c}$, for the two-dimensional Ising model with a double-blocked
  interaction profile.  $\tilde{\chi}=\chi'/C(R)$, where $C(R)$ is a
  range-dependent correction factor that accounts for the fact that the
  critical amplitude for small~$R$ deviates from the asymptotic range
  dependence. This only introduces a shift along the vertical axis. The solid
  curve indicates the crossover function for the block-shaped interaction
  profile and the dashed lines mark the mean-field (``MF'') and the Ising
  asymptote. The numbers in the key refer to the values for $R_1^2$ and~$R_2^2$
  for each interaction profile.}
  \label{fig:con-dblk}
\end{figure}
The data for the susceptibility are shown in Fig.~\ref{fig:con-dblk}. Since the
critical amplitude of the susceptibility varies as $R^{-3/2}$, a data collapse
is obtained for $\chi'/R^2$. One observes how all data perfectly coincide with
the crossover curve for the equivalent-neighbor model. The deviations from the
curve at the right-hand side of the graph are caused by the fact that,
sufficiently close to~$T_{\rm c}$, the diverging correlation length is
truncated by the finite system size.  For the systems with $R_1^2=4$
and~$R_1^2=93$ (both with $R_2^2=140$) this happens in the figure at different
temperatures, despite the fact that they have very similar values for the
effective range~$R$. The reason for this is that the data points pertain to
different system sizes, viz.\ $L=1000$ and~$L=300$, respectively. The inset
shows that for the \emph{same} system size ($L=300$) the data points for both
systems virtually coincide, even in the finite-size regime.  The crossover
function for the connected susceptibility appears to be insensitive to the
introduction of an additional length scale in the interaction profile.  We
hence view this graph as a strong indication that crossover functions possess a
considerable degree of universality.

\newpage
\section{Outlook and Conclusion}
We have discussed a cluster algorithm for spin models with long-range
interactions that is several orders of magnitude more efficient than
conventional algorithms. Its usefulness has been demonstrated for a system with
power-law interactions as well as for systems with medium-range interactions.
In both cases, physical results have been obtained that until now could not be
generated within a reasonable amount of computing time. However, the scope of
the algorithm goes far beyond what could be presented here. As far as purely
ferromagnetic interactions are concerned, it can be efficiently applied to any
interaction profile, including anisotropic ones. Also the order parameter must
not necessarily be Ising-like: the extension to general $\mathrm{O}(n)$
models~\cite{lr-alg} is similar to that for the original Wolff algorithm and
the first application to a Potts model has already been
published~\cite{glumac98}.

The generalization to a situation in which additional antiferromagnetic
interactions are present is, in principle, also possible. Whereas competing
interactions will move the system away from the percolation threshold (and thus
critical slowing down will no longer be optimally suppressed), one still has
the advantage that not every individual spin has to be considered for inclusion
in the cluster. Finally, it would be interesting to see whether this algorithm
can be useful in other fields of physics where cluster algorithms have come to
flourish.

\section*{Acknowledgements}
I wish to thank Henk Bl\"ote and Kurt Binder for the fruitful collaboration
that has led to the work presented in this paper. I am grateful to Holger
Me{\ss}ingfeld for producing the data shown in Fig.~\ref{fig:psi} and
performing the least-squares fits for the inverse-square chain.



\end{document}